\documentclass[twocolumn,amsmath,amssymb,prl,superscriptaddress,nofootinbib]{revtex4}
\usepackage{graphicx}
\usepackage{dcolumn}
\usepackage{bm}
\usepackage{booktabs}

\begin{document}

\title{Self-accelerated universe induced by repulsive effects as an alternative to dark energy and modified gravities}

\author{Orlando Luongo}
\affiliation{Dipartimento di Fisica, Universit\`a di Napoli ''Federico II'', Via Cinthia, I-80126, Napoli, Italy.}
\affiliation{Istituto Nazionale di Fisica Nucleare (INFN), Sez. di Napoli, Via Cinthia, I-80126, Napoli, Italy.}
\affiliation{Instituto de Ciencias Nucleares, Universidad Nacional Aut\'onoma de M\'exico, AP 70543, M\'exico DF 04510, Mexico}

\author{Hernando Quevedo}
\affiliation{Instituto de Ciencias Nucleares, Universidad Nacional Aut\'onoma de M\'exico, AP 70543, M\'exico DF 04510, Mexico}
\affiliation{Dipartimento  di Fisica and ICRA, ``Sapienza" Universit\`a di Roma, Piazzale Aldo Moro, I-00185, Roma, Italy.}

\begin{abstract}
The existence of current-time universe's acceleration is usually modeled by means of two main strategies. The first makes use of a dark energy barotropic fluid entering \emph{by hand} the energy-momentum tensor of Einstein's theory. The second lies on extending the Hilbert-Einstein action giving rise to the class of extended theories of gravity. In this work, we propose a third approach, derived as an intrinsic geometrical effect of space-time, which provides repulsive regions under certain circumstances. We demonstrate that the effects of repulsive gravity naturally emerge in the field of a homogeneous and isotropic universe. To this end, we use an invariant definition of repulsive gravity based upon the behavior of the curvature eigenvalues. Moreover, we show that repulsive gravity counterbalances the standard gravitational attraction influencing both late and early times of the universe evolution. This phenomenon leads to the present speed up and to the fast expansion due to the inflationary epoch. In so doing, we are able to unify both dark energy and inflation in a single scheme, showing that the universe changes its dynamics when ${\ddot H\over H}=-2\dot H$, at the repulsion onset time where this condition is satisfied. Further, we argue that the spatial scalar curvature can be taken as vanishing  because it does not affect at all the emergence of repulsive gravity.  We check the goodness of our approach through two cosmological fits involving the most recent union 2.1 supernova compilation.
\end{abstract}


\date{\today}

\maketitle



\paragraph{{\bf \emph{Introduction}}}
General relativity predicts a universe characterized by a decelerated expansion, if only dark matter and baryons are involved within the energy-momentum tensor. Observations, however, predict a late-time accelerated universe, pushed up by some sort of exotic constituent whose nature is today unknown \cite{rev1}.
To account for this experimental fact, cosmologists employ in general two different strategies. The first one consists in extending the Hilbert-Einstein action, modifying general relativity by means of additional degrees of freedom \cite{rev2}. The second one takes into account an additional fluid, dubbed \emph{dark energy}, which enters  Einstein's equations and represents a source for the cosmic speed up \cite{rev3}. Both those treatments are based on the hypothesis that space-time geometry is \emph{fueled} by components entering the energy-momentum tensor $T_{\mu\nu}$. In such a way, the energy-momentum tensor becomes the source for gravitation and modifies the geometry itself. However, if we rewrite Einstein's equations in an equivalent way  as   $T_{\mu\nu}-\frac{1}{2}g_{\mu\nu}T=\chi R_{\mu\nu}$, it is possible to reverse the interpretation of $T_{\mu\nu}$. Indeed, one can address a dynamical problem by considering an energy-momentum tensor induced by geometry, and not {\it vice-versa}. The philosophy is completely different from the one used in higher-order theories of gravity, since no extensions are required, but only a physical mechanism inside the Einstein equations themselves, providing a \emph{geometrical source term}.

It is therefore possible to imagine that under certain circumstances the geometry could correspond to a repulsive field that counterbalances the attraction of gravity and induces an accelerated universe.  This \emph{third way} of handling the universe dynamics becomes a natural consequence of general relativity itself, instead of a new additional ingredient that must be added into the energy-momentum tensor. Indications of repulsive effects in gravity are not new, specially in the case of naked singularities \cite{divsic}. Several intuitive approaches have been proposed to understand repulsive gravity, but only recently an invariant definition in terms of geometric objects was formulated \cite{ioeherny}. Indeed, it was shown that the behavior of the curvature tensor eigenvalues can be used to find the places where repulsive gravity appears. This was explicitly shown in the case of naked singularities with black hole counterparts where it was found that there exists a repulsion region which is always situated very close to the source of gravity.

Here, we extend this approach to the case of a homogeneous and isotropic universe, modeled by the Friedmann-Robertson-Walker (FRW) metric. We consider the effects of repulsive gravity as a source for the cosmic speed up, and find the stage at which the universe starts accelerating, corresponding to a phase in which repulsive gravity dominates over attraction. We therefore interpret the acceleration of the universe in terms of a natural outcome due to the non-linearity of large-scale cosmology which is reflected in the eigenvalues of the curvature tensor. We characterize gravitational repulsion by evaluating precise values for the observable acceleration parameter $q$ and its variation $j$. We also find that the universe dynamics does not depend upon the scalar curvature; interestingly, this could be interpreted as an explanation of the experimental evidence for that the universe appears spatially flat at all stages of its evolution, and enables us to unify the inflationary phase with the accelerated phase in a single description. Finally, we find  an approximate effective density induced by repulsive gravity, assuming that the conditions valid at the moment of the repulsion onset continue holding at all times. We check the goodness of our approach through a numerical test based on supernova union 2.1 data, performed on a grid in which all cosmic parameters evolve. Particularly, we employ two hierarchical experimental tests. In the first one, we take all coefficients free to span in allowed prior domains. In the second one, we fix the acceleration parameter by means of the most recent Planck results. We definitively show compatible limits between our theoretical predictions and numerical outcomes.


\paragraph{{\bf \emph{Repulsive effects in a homogeneous and isotropic universe}}}

Although there are several intuitive ways to define repulsive gravity, only recently an invariant definition  in terms of the curvature tensor eigenvalues was formulated  which gives physically reasonable results in the case of all naked singularities investigated so far \cite{ioeherny}. The idea is that the curvature tensor eigenvalues contain in an invariant manner all the information about the behavior of gravity. To this end, it is convenient to introduce an orthonormal frame
$\vartheta^a$ which represents the simplest choice for an observer to perform local time, space and gravity measurements so that all the quantities related to this frame are coordinate independent. The orthonormal tetrad is determined by the relationships $ds^2 = g_{\mu\nu} dx^\mu dx^\nu= \eta_{ab}\vartheta^a\otimes\vartheta^b$, with $\eta_{ab}={\rm diag}(-1,1,1,1)$, and $\vartheta^a = e^a_{\ \mu}dx^\mu$. Furthermore, the first and second Cartan equations enables one to compute explicitly the Riemann curvature components in this frame.

For the analysis of the eigenvalues it is convenient to consider the bivector representation of the Riemann tensor, which is also useful for finding its irreducible representation with respect to the Lorentz group. We use the notations and conventions introduced in \cite{hern}, according to which each bivector index
$A,B=1,\ldots,6$  corresponds to two tetrad indices, i.e., $A\rightarrow ab$, with the convention
$1\rightarrow 01,\ 2\rightarrow 02,\ 3\rightarrow 03,\  4 \rightarrow 23, \  5\rightarrow 31, \  6\rightarrow 12$.
Then, the curvature tensor can be expressed as a $(6\times 6)-$matrix which, in the case of the FRW orthonormal frame
$\vartheta^0 = dt,\
\vartheta^1 = a(t) dr/\sqrt{1-kr^2},\
\vartheta^2 = a(t) r d\theta,\
\vartheta^3 = a(t) r \sin\theta d\varphi$, reduces to the diagonal matrix
\begin{equation}
\mathcal R = {\rm diag}(\lambda,\lambda,\lambda,\tau,\tau,\tau)\ ,
\end{equation}
\begin{equation}
\lambda = -\frac{\ddot a}{a}= \frac{1}{6} (3 P + \rho) \ ,\quad \tau = \frac{\dot a^2 + k}{a^2} = \frac{1}{3} \rho \ ,
\label{ev}
\end{equation}
where we have used the Friedmann equations and $8\pi G=1$. It then follows that the only independent eigenvalues of the curvature tensor are $\lambda$ and $\tau$.

According to our invariant definition of repulsive gravity \cite{ioeherny}, the curvature eigenvalues reflect the behavior of the gravitational interaction and
if gravity becomes repulsive in some
regions, the eigenvalues must change accordingly; for instance, if repulsive gravity becomes
dominant at a particular point, one would expect at that point a
change in the sign of at least one eigenvalue.  Moreover, if the gravitational field does not diverge at infinity, the eigenvalue must have an extremal at some point before it changes its sign. This means that the extremal of the eigenvalue can be interpreted as the onset of repulsion.

Let us now consider the eigenvalues of the FRW curvature tensor. First, we note that $\tau$ is positive definite by virtue of the energy conditions and so it does
not contain any information about repulsive gravity. It therefore represents a constraint eigenvalue which fixes the total energy budget pushing up the universe, but with no information on its evolution.

Instead, $\lambda$ can change its sign in the interval where the pressure $P$ is negative and, so, the universe can accelerate.  Thus, $\lambda$ represents the dynamical eigenvalue which dominates the spacetime dynamics, predicting the acceleration. Moreover, $\lambda$ does not depend explicitly on the value of the constant $k$ which is the first interesting consequence of our approach. Indeed, this means that the universe does not need scalar curvature to accelerate so that the curvature density
$\Omega_k\equiv -\frac{k}{a^2H_0^2}$ can be taken arbitrarily small, without affecting the universe dynamics. This theoretical result is in agreement with observations \cite{planck}. On the other hand, this could be considered as a theoretical explanation of the spatial flatness of the universe.

\begin{figure}[ht]
\includegraphics[scale=0.65]{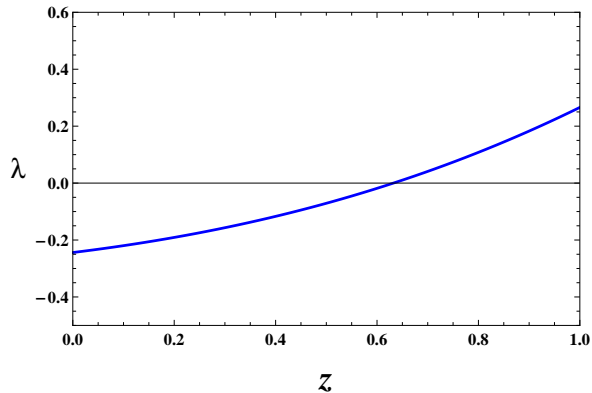}
\caption{{\it{Plot of $\lambda$ in the redshift domain $z\in[0;1]$ with $H$ and $q$ from the $\Lambda$CDM paradigm with $\Omega_m=0.318$ and $h_0=0.68$.}}}
\label{figura2}
\end{figure}

To further analyze the repulsive effects in the dynamics of our the universe, we introduce the acceleration parameter
\cite{orlcosmo}
\begin{equation}\label{q}
q= - \frac{\ddot a}{aH^2} =-1-\frac{\dot H}{H^2}\,,
\end{equation}
to obtain
\begin{equation}\label{lambda4}
\lambda=qH^2\,
\end{equation}
so that  $\lambda<0$ when the universe is accelerating, whereas for $\lambda>0$ the universe decelerates.
The condition
\begin{equation}
\dot H=-H^2
\label{accond}
\end{equation}
determines the special point $\lambda=0$ at which the acceleration parameter vanishes and the universe undergoes a transition from
a decelerating phase into an accelerating phase. The particular time $t=t_{acc}$ at which this phase transition occurs follows from the integration
of Eq. (\ref{accond}), i.e.,
\begin{equation}
H_{acc} = H(t_{acc}) = \frac{1}{t_{acc} + c_0}
\label{tacc}
\end{equation}
where $c_0$ is an integration constant. The behavior of the eigenvalue $\lambda$ is illustrated in
Fig. \ref{figura2} for $q$ and $H$ of the $\Lambda$CDM model.

As mentioned above, the onset of repulsion is determined by an extremal of the eigenvalue, i.e. $\dot \lambda =\dot q H^2 + 2q H \dot H= 0$. Then, using Eq. \eqref{q},
we obtain the onset condition
\begin{equation}\label{versotra}
\dot q = 2H(1+q)q\,.
\end{equation}
Using the identity \cite{relazione}
\begin{equation}
 \dot q  = - ( j-q-2q^2) H  \ ,
\end{equation}
where the jerk parameter $j = {\dddot a}/({a H^3})$ determines the change of the acceleration \cite{ioancora}, the repulsion onset condition (\ref{versotra}) reduces to
\begin{equation}
j = - q \ ,
\label{repcond}
\end{equation}
implying that at that moment of time $t=t_{rep}$ the jerk must be positive because the onset happens when the universe is still decelerating $(q>0)$. By using the definitions
of the cosmographic parameters, we can rewrite (\ref{repcond}) as
\begin{equation}\label{ghhh}
\frac{\ddot H}{H}=-2\dot H\,,
\end{equation}
a relationship that can be integrated and yields
\begin{equation}
H_{rep}= H(t_{rep}) = \alpha\tanh[\alpha\left(t_{rep}+\beta\right)]\,,
\label{hrep}
\end{equation}
where $\alpha$ and $\beta$ are integration constants. This is the behavior of the Hubble parameter at the onset of repulsion. From that point on, the value of the acceleration parameter $q$ diminishes until it vanishes at $t=t_{acc}$ (cf. Fig. \ref{figura2}). It seems reasonable to construct a cosmological model based
upon the behavior of the Hubble parameter near the repulsion onset point (\ref{hrep}), where the acceleration begins to play an important role in the universe dynamics. In fact, one expects acceleration to be important in late and early phases of the universe evolution.
Hence, we can assume that Eq.(\ref{hrep}) is valid from the transition time up to $z\rightarrow0$ and also  during the inflationary phase.

As the redshift decreases, corresponding to the limit of $t\rightarrow t_0$, with $t_0$ our present time, we get
\begin{subequations}\label{approximatiotolambda}
\begin{align}
\alpha\tanh[\alpha\left(t_0+\beta\right)]&\approx \rho_\Lambda\,,\\
\alpha^2\cosh^{-2}[\alpha\left(t_0+\beta\right)]&\approx-3H_0\rho_m-\dot \rho_m\,,
\end{align}
\end{subequations}
where we considered $\rho_\Lambda\equiv3\Lambda$ the cosmological constant density in units in which $8\pi G=1$, $\rho_m$ is the today matter density and, furthermore, we used the continuity equation $\dot \rho+3H(P+\rho)=0$ for the case of a late-time universe, in which $P\sim P_\Lambda$ and $\rho\sim\rho_m+\rho_\Lambda$. The above equations fix limits over $\alpha$ and $\beta$ at late times.

 On the other hand, at the inflationary phase we have
\begin{subequations}\label{approximatiotoinfla}
\begin{align}
\alpha\tanh[\alpha\left(t_{i}+\beta\right)]&\approx V(\phi)\,,\\
\alpha^2\cosh^{-2}[\alpha\left(t_i+\beta\right)]&\approx-\sqrt{3\,V(\phi)}\dot\phi\,,
\end{align}
\end{subequations}
where we considered the relation $\rho\approx\rho_\phi=\frac{1}{2}\dot \phi+V(\phi)$ and $P\approx P_\phi=\frac{1}{2}\dot \phi-V(\phi)$, with $\phi$ and $V(\phi)$ the field and potential determining the inflation, respectively. We also took into account the slow-roll approximation $V(\phi)\gg \dot \phi$ and $t_i$, the time at which inflation occurred. Those relations impose bounds on $\alpha$ and $\beta$ at the times in which the universe accelerated.
In other words, the mechanism of acceleration is induced for different epochs in the same way, but with different coupling constants, whose values are associated to the typology of the acceleration involved, i.e., with  dark energy or inflation.


\paragraph{{\bf \emph{Numerical outcomes from experimental fittings}}}

We carry out two analyses based on supernova standard candles, employing the most recent union 2.1 compilation. This survey has been built up by means of 580 measurements of supernova magnitudes. To each magnitude $\mu$, one associates the corresponding redshift $z$ and magnitude error $\delta\mu$, assuming the redshift error to be negligible.

Under the hypothesis of a perfect Gaussian distribution, it is possible to relate the luminosity distance $D_L$ to the magnitude $\mu$. Afterwards, since $D_L=D_L(H_0,q_0,j_0)$, one gets:
\begin{eqnarray}\label{dLTaylor}
    D_L = \frac{z}{H_0}\,\left(1 + d_{L1}z + d_{L2}z^2 + \ldots\right)\,,
\end{eqnarray}
with the coefficients $d_{L1}=\frac{1}{2}(1-q_0)$ and $d_{L2}=\frac{1}{6}(-1-j_0+q_0+3q_0^2)$ \cite{orlpade}. Notice that we \emph{a priori} assumed $\Omega_k=0$, without losing generality, since $\lambda$ does not depend upon it.

We therefore consider a parameter-fitting procedure which makes use of a grid in which all parameters evolve, in their corresponding priors, i.e. $H_0\in[60,85]$, $q_0\in[-1,0]$, $j_0\in[-2,2]$. We minimize the chi-square function:
\begin{equation}\label{chiqua}
\chi^2 = \mathcal A - \frac{\mathcal B^2}{\mathcal C} + \mathcal D\,,
\end{equation}
where $ \mathcal A = {\bf x}^T C^{-1} {\bf x}$, $\mathcal B = \sum_i (C^{-1}{\bf x})_i$, $\mathcal C = \text{Tr}[\,C^{-1}\,]$ and $\mathcal D = \log \left( \frac{\mathcal C}{2\pi}\right)$ and we get the results provided in Tab. \ref{tabella}.
\begin{figure}[htbp]
\centering
\includegraphics[width=80mm]{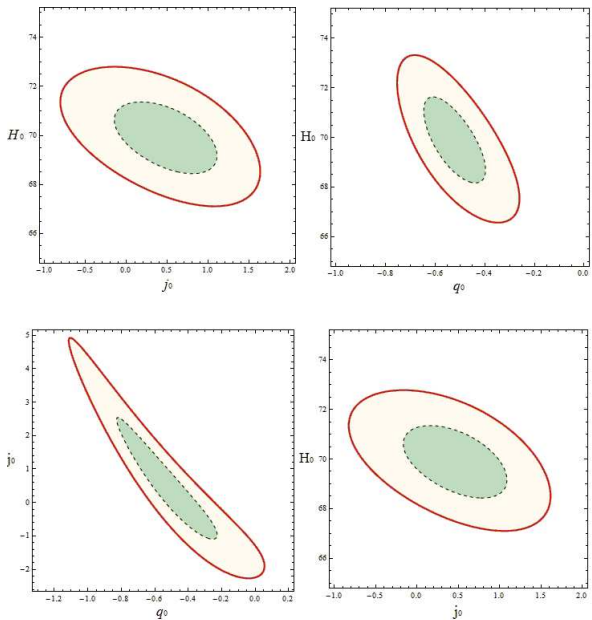}

\caption{\label{contorni} {\it{Contours for the experimental results. We plot the graphics of $j_0$ versus $H_0$ (top left figure), $q_0$ versus $H_0$ 
(top right), $q_0$ versus $j_0$ (button left) and $j_0$ versus $H_0$ of the second fit (button right), in which $q_0$ is fixed in the framework of the $\Lambda$CDM model, with $\Omega_m=0.318$, compatible with modern Planck results. In the contour plots, we indicate $1\sigma$ (inside ellipse) and $2\sigma$ (outer ellipse) with different colors, corresponding to  $68\%$ and $95\%$ confidence levels. }}}
\end{figure}
The advantage of using the union 2.1 compilation on a grid is that previous systematics of other catalogs have been reduced. Thus, it is easy to show that the {\it $\chi$-squared} function \eqref{chiqua} permits to maximize the likelihood function:
\begin{equation}\label{lik}
    \mathcal{L}
\propto \exp (-\chi^2/2 )\,,
\end{equation}
with a good approximation.

In Tab. \ref{tabella}, we present the results derived from our experimental analysis. Fig. \ref{contorni} contains the corresponding contour plots with the 
$1\sigma$ and $2\sigma$ of our coefficients. The values of $q_0$ and $j_0$ are basically close to each other, as theoretically predicted. Unfortunately, the higher errors on $j_0$ do not allow to characterize this property at a $5\sigma$ confidence level which would definitively suggest the validity of our model. However, incoming refined data are expected to certify this experimental fact, to see how the coefficients $q_0$ and $j_0$ differ from each other.

\begin{center}
\begin{table}[ht]
\begin{tabular}{c|c|c}
\hline\hline

{\small $\quad$ Parameter $\quad$}  &   {\small $\qquad$ Fit1 $\qquad$ }    &
{\small $\qquad$ Fit2 $\qquad$ }\\

\hline

{\small$\chi^2$}       & {\small $ 0.975$}        &          {\small $0.976$}\\[0.8ex]

{\small$H_0$}       & {\small $ 69.887$}{\tiny ${}_{ -1.253}^{ +1.276}$}        &{\small $69.873$}{\tiny ${}_{-1.253}^{+2.528}$}\\[0.8ex]

{\small$q_0$}       & {\small $-0.527$}{\tiny ${}_{-0.088}^{+0.093}$}      & {\small$-0.523$}{\tiny ${}_{---}^{---}$}\\[0.8ex]

{\small$j_0$}       & {\small $0.501$}{\tiny ${}_{-0.527}^{ +0.558}$}     & {\small $0.482$}{\tiny ${}_{-0.527}^{+0.558}$}\\[0.8ex]

{\small$\lambda$}       & {\small $-0.257$}{\tiny ${}_{-0.052}^{+0.055}$}     & {\small $-0.255$}{\tiny ${}_{-0.009}^{+0.018}$}\\[0.8ex]

\hline \hline
\end{tabular}
\caption{\label{tabella} {\it{Table of our experimental results and numerical limits. We enumerate the coefficients $q_0$ and $j_0$, showing that their absolute values are surprisingly close to each other, as predicted by our repulsion model in a FRW universe. We also evaluate $\lambda$ at present time, through the logarithmic propagation of experimental errors. The subscript $0$ indicates the corresponding quantity is evaluated at present time. The fits are two: in the first one, we leave all coefficients free to vary, whereas in the second one we assume a fixed $q_0$ from the $\Lambda$CDM calibration, as suggested by Planck measurements.}}}
\end{table}
\end{center}


\paragraph{{\bf \emph{Conclusions and perspectives}}}
We presented a new approach to describe the universe dynamics at current and early phases of its evolution. We showed that it is possible to infer a process enabling the universe to accelerate by means of repulsive effects. To investigate such a phenomenon, we use an invariant definition of repulsive gravity in terms of the curvature tensor eigenvalues and we got two independent eigenvalues: $\lambda$ and $\tau$. It turns out that the dynamics of the repulsive component is governed by a particular eigenvalue $\lambda$ whose behavior at the moment of the repulsion onset, $\dot \lambda =0$, allows us to construct a cosmological model based upon the behavior of the Hubble parameter at that moment.  Moreover, the repulsion onset condition implies that the cosmographic parameters must satisfy the algebraic relationship  $j+q=0$, which is then shown to be in good agreement with observations.  As a result of this analysis, we were able to feature an \emph{effective emergent dark energy fluid}, which depends on two free integration constants whose values are fixed for dark energy and inflationary phases. In so doing, we demonstrated that this process fuels the universe acceleration without the need of scalar curvature $\Omega_k$. We therefore found a theoretical explanation for assuming a spatially flat universe at both late and early phases of its evolution. For these reasons, we believe our treatment to represent a  \emph{third way} to account the problem of cosmic speeding up, instead of considering barotropic dark energy fluids or modifications of general relativity. Finally, we checked the goodness of our cosmological approach by fitting the union 2.1 compilation, inferring cosmic bounds on the observable quantities and showing that our picture is compatible with present time observations. Future developments will better characterize our model at different stages of the universe expansion history, to better clarify how eigenvalues change sign. Additional efforts will be also spent to better constrain the cosmological parameters of interest with different data sets and combined data surveys.\\


\section*{Acknowledgements}
We wish to thank Dr. Andrea Geralico and Dr. Damiano Tommasini for the support with the numerical analysis and Prof. Salvatore Capozziello for useful discussions. This work was partially supported by DGAPA-UNAM, Grant No. 113514, and CONACyT, Grant No. 166391.

\end{document}